# Effect of PVA doping on flux pinning in Bulk $MgB_2$

**Arpita Vajpayee[1], V. P. S. Awana[1,*], S. Balamurugan[2], E. Takayama-Muromachi[2] H. Kishan[1], G. L. Bhalla[3]**

[1]National Physical Laboratory, K. S. Krishnan marg, New Delhi, India

[2] National Institute for Materials Science (NIMS), 1-1 Namiki, Tsukuba, Ibaraki 305-0044, Japan

[3]Dept. of Physics and Astrophysics, Univ. of Delhi, New Delhi, India

**Abstract**

The synthesis and characterization of PVA (Poly Vinyl Acetate) doped bulk $MgB_2$ superconductor is reported here. PVA is used as a Carbon source. PVA doping effects made two distinguishable contributions: first enhancement of $J_c$ field performance and second an increase in $H_{c2}$ value, both because of carbon incorporation into $MgB_2$ crystal lattice. The susceptibility measurement reveals that $T_c$ decreased from 37 to 36 K. Lattice parameter 'a' decreased from 3.085Å to 3.081Å due to the partial substitution of Carbon at Boron site. PVA doped sample exhibited the $J_c$ values greater than $10^5$ A/cm$^2$ at 5 & 10 K at low fields; which is almost 3 times higher than the pure one, while at high fields the $J_c$ is increased by an order of magnitude in comparison to pure $MgB_2$. From $\rho(T)H$ measurements we found higher $T_c$ values under magnetic field for doped sample; indicating an increase in $H_{c2}$. Also the magnetization measurements exhibited a significant enhancement in $H_{irr}$ value. The improved performance of PVA doped $MgB_2$ can be attributed to the substitution of carbon at boron site in parent $MgB_2$ and the resulting impact on the carrier density and impurity scattering. The improved flux pinning behavior could easily be seen from reduced flux pinning force plots.

Key Words: $MgB_2$ superconductor, magnetization and Flux pinning

*: E-mail- awana@mail.nplindia.ernet.in

## Introduction

After the discovery of superconductivity [1] in $MgB_2$, enormous efforts had been directed towards making this material technologically important for applications [2-7]. For practical applications of a superconductors, besides its superconducting transition temperature ($T_c$), further crucially important parameters are critical current density at higher fields $J_c(H)$ and irreversibility field $H_{irr}$. The $J_c$ of pristine $MgB_2$ drops rapidly in high magnetic fields due to weak pinning centers and low upper critical field $H_{c2}$. Chemical doping is generally thought of as an effective approach to improve the superconducting performance of $MgB_2$. Many of the groups have doped several elements and *nano-* particles in $MgB_2$ [2-7] to improve its superconducting performance via effective pinning of flux vortices. Carbon doping had given good results in this aspect [8-11]. The solubility of Carbon in $MgB_2$ lattice varies from say 1.25% to 30% [12, 13]. Mostly Carbon is doped in $MgB_2$ by direct reaction of Magnesium, Boron and Carbon powders [12-15]. Carbon is the only element which partially substitutes at Boron sites into the $MgB_2$ lattice. The B-site partial substitution of C creates disorder in the lattice and hence enhances the $H_{c2}$ and $J_c(H)$ via intrinsic pining. Carbon has one more electron than Boron, so it is expected that Carbon doping would modify the superconductivity in $MgB_2$. We explored here this role of Carbon, but we didn't put the Carbon directly in $MgB_2$ and rather added the PVA solution in $MgB_2$ as a Carbon source. This way one produces fresh Carbon at the time of formation of $MgB_2$. In most of the reports, dopants were introduced through dry mixing process that might be a cause of inhomogeneous distribution of dopants with in the matrix material. To avoid this we used the Carbon source in liquid form as PVA (Poly Vinyl Acetate) solution. The Carbon atoms in present case may either substitute at Boron site in lattice or remain as additives in host matrix. Both ways (substitution/addition) Carbon in $MgB_2$ is supposed to enhance the $J_c(H)$ and irreversibility field $H_{irr}$ via intrinsic/extrinsic pinning of flux lines. We report in this article the results of $Hc_2$, $J_c(H)$ and $H_{irr}$ for pristine and PVA added $MgB_2$, and found that all these physical parameters improve dramatically with PVA addition.

## Experimental

The PVA doped $MgB_2$ sample is prepared by solid-state reaction route. The constituent powders (Magnesium and Boron) were well mixed in stoichiometric ratio through grinding for 1.5 hour. We dissolved half gram of PVA (Poly Vinyl Acetate) powder in 5 ml of Acetone. Then we added 3ml. of this PVA solution in 2 grams of $MgB_2$ raw powder, which was followed by again grinding in order to form the homogenous mixture. The mixture was palletized using hydraulic press. The pallets were enclosed in soft iron tube and that was put in tubular furnace at 850$^o$C in argon atmosphere. The heating rate was about 425$^o$C per hour and the holding time was 2.5 hours. After this annealing treatment the furnace was allowed to cool down naturally. The x-ray diffraction pattern of compound was taken using $CuK_\alpha$ radiation. The magnetoresistivity, $\rho(T)H$, was measured with $H$ applied perpendicular to current direction, using four-probe technique on *Quantum Design PPMS*. The magnetization measurements were carried out using *Quantum Design MPMS-XL.*

## Results and Discussion

Fig. 1 shows the XRD patterns of pure $MgB_2$ and PVA doped $MgB_2$ samples. This figure depicts that both the samples are having hexagonal pure $MgB_2$ phase with traces of a little amount of MgO impurity. The lattice parameters for pristine and PVA added $MgB_2$ compounds are $a$ = 3.085 Å, $c$ = 3.520 Å, and $a$ = 3.081 Å, $c$ = 3.521 Å respectively. The $c/a$ value of the pure compound is close to 1.142, which is known to be optimum for stoichiometric $MgB_2$. The reduction in 'a' parameter is due to the substitution of Carbon for Boron in the lattice of $MgB_2$ [14, 15], which is released from the heating of added PVA during synthesis. On the other hand lattice parameter 'c' is slightly increased. The variation of lattice parameters clearly indicate that Carbon is substituted at Boron site in PVA added $MgB_2$ compound.

Resistivity vs temperature plots under various applied magnetic fields $\rho(T)H$ are shown in fig. 2(a) and 2(b) for the undoped and doped sample respectively. The residual resistivity ratio (RRR = $R_{T300K}/R_{Tonset}$) for pure sample and PVA added samples are 2.36 and 2.24 respectively. The higher values of room temperature resistivity for doped sample indicate that the impurity scattering is stronger due to the Carbon substitution at Boron sites. The superconducting transition temperature $T_c$ ($\rho$ =0) for pure $MgB_2$ is ≈ 37 K without applying any field and is decreased to 18K under applied field of 8 Tesla (This is shown in inset of fig. 2(a)). As we added PVA into pure $MgB_2$ the $T_c$ is decreased to 36K due to carbon doping. This was the situation at zero Tesla field while at higher fields the superconducting performance of doped sample is improved. For example the $T_c$ ($\rho$ =0) is observed at 20K under 9 Tesla applied field for the PVA added sample (inset of fig. 2(b)). This effect was attributed to the role of carbon as trapping centers in the presence of magnetic fields.

Susceptibility vs temperature $\chi(T)$ plot for doped and undoped samples in field-cooled and zero field-cooled cases is shown in fig. 3. Both samples exhibit sharp one step superconducting transition. The superconducting transition temperature $T_c$ decreased with PVA doping. This is because of partial substitution of Carbon for Boron in $MgB_2$, resulting in a decrease of the carrier (hole) concentration and consequently a reduction in the $T_c$ [13, 16]. The amplitude of diamagnetic signal is changed slightly by PVA doping in zero- field cooled case. Though the pinning is more in PVA doped sample, the weak field- cooled transitions for both samples indicate strong pinning in them. The volume fraction of superconductivity is not estimated, because of weak field-cooled transitions.

Fig. 4 shows the magnetization plots for undoped and doped samples at 5 K. The magnetization loop for the undoped sample is relatively narrow and small, revealing comparatively low irreversibility field. On the other hand there is clear indication of improvement in irreversibility field $H_{irr}$ for doped sample because its $M(H)$ loop is wide and quite open even at 7 Tesla.

Using width of magnetization plots we calculated the magnetic critical current density ($J_c$) for both samples invoking Bean's critical state model. The dependence of critical current density on applied magnetic field $J_c(H)$ at 5, 10 & 20 K is shown in fig. 5. The $J_c$ value for doped sample under 6 Tesla applied field reached at $0.93 \times 10^4$ A/cm$^2$ and $4.49\times 10^3$ A/cm$^2$ at 5 and 10 K respectively. These values are an order of magnitude higher than as for the pure $MgB_2$. This is an indication of increased pinning strength in the samples at higher fields.

To confirm the improved flux pinning behavior through PVA doping, we show in fig. 6 the plots of the field dependence of normalized flux pinning force [17, 18] at 5 & 10 K. It is clear from this figure that the pinning force for doped sample is much higher than the pure $MgB_2$ above 1.5 Tesla, indicating enhanced flux pinning force in high fields. Further the peak pinning force is shifted towards higher fields for the PVA added sample at all studied temperatures.

The temperature dependence of upper critical field $H_{c2}$ for both samples is shown in fig.7. There is a clear shift in upper direction in $H_{c2}$ value for doped sample in comparison to pure sample. The carbon substitution into boron site in lattice is responsible for the enhancement in $H_{c2}$ value. The Two-band impurity scattering model of charge carriers had indicated increased intra-band scattering via shortening of the electron mean free path, which could be a possible reason for increased $H_{c2}$ value [19].

## Conclusions

In summary, polycrystalline C doped $MgB_2$ samples are prepared with PVA being used as the carbon source material. Though $T_c$ decrease slightly significant improvement is seen in $J_c(H)$, $H_{irr}$ and $H_{c2}$ performance. In present study C substitution and $MgB_2$ formation took place simultaneously. PVA provided fresh and reactive Carbon at the time of reaction because of its decomposition just before the formation temperature of $MgB_2$.

The incorporation of Carbon atoms into the $MgB_2$ crystal results in distortion of lattice and in-turn the improved $J_c(H)$ behavior. Briefly, the addition of PVA in $MgB_2$ had resulted in excellent superconducting performance due to the incorporation of Carbon at Boron site in pristine compound.

**Acknowledgement**

The authors from NPL would like to thank Dr. Vikram Kumar (DNPL) for his great interest in present work.

**Figure captions**

Figure 1. The x-ray diffraction pattern for Pure and doped $MgB_2$

Figure 2. Resistivity vs temperature plots under various applied magnetic field for (a) Pure $MgB_2$ and (b) PVA doped $MgB_2$

Figure 3. Temperature dependence of susceptibility in zero-field cooled and field cooled case for doped and undoped samples

Figure 4. Magnetization loop $M(H)$ at 5K for pure and PVA doped $MgB_2$

Figure 5. Critical current density ($J_c$) variation with respect to applied magnetic field ($H$) at different temperature 5,10 & 20K

Figure 6. Reduced flux pinning force ($F_p/F_{p,max}$) variation with magnetic field for doped and undoped samples at 5K & 10K

Figure 7. Upper critical field ($H_{c2}$) as a function of temperature for pure and PVA doped $MgB_2$

Fig. 1

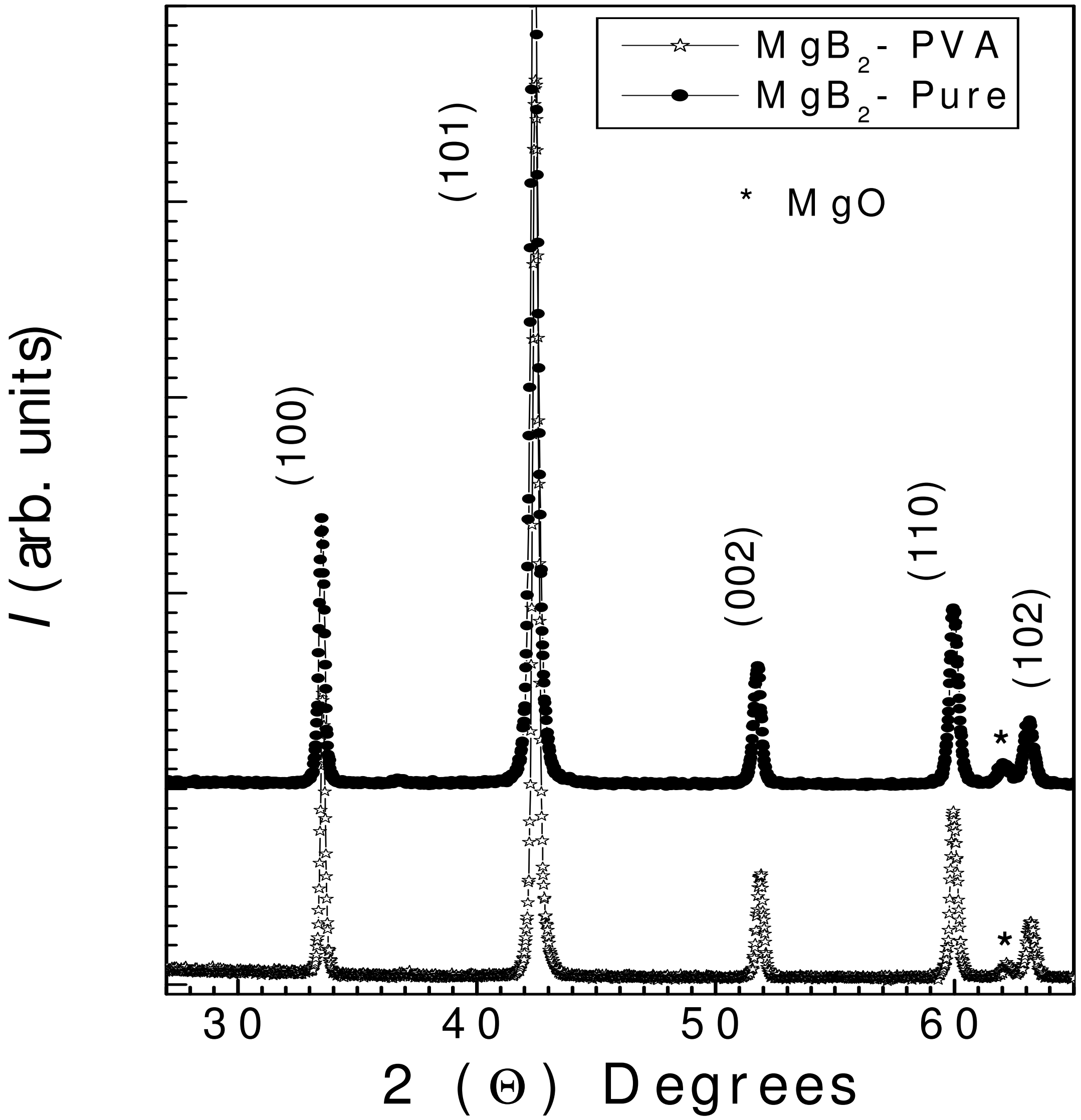

Fig. 2

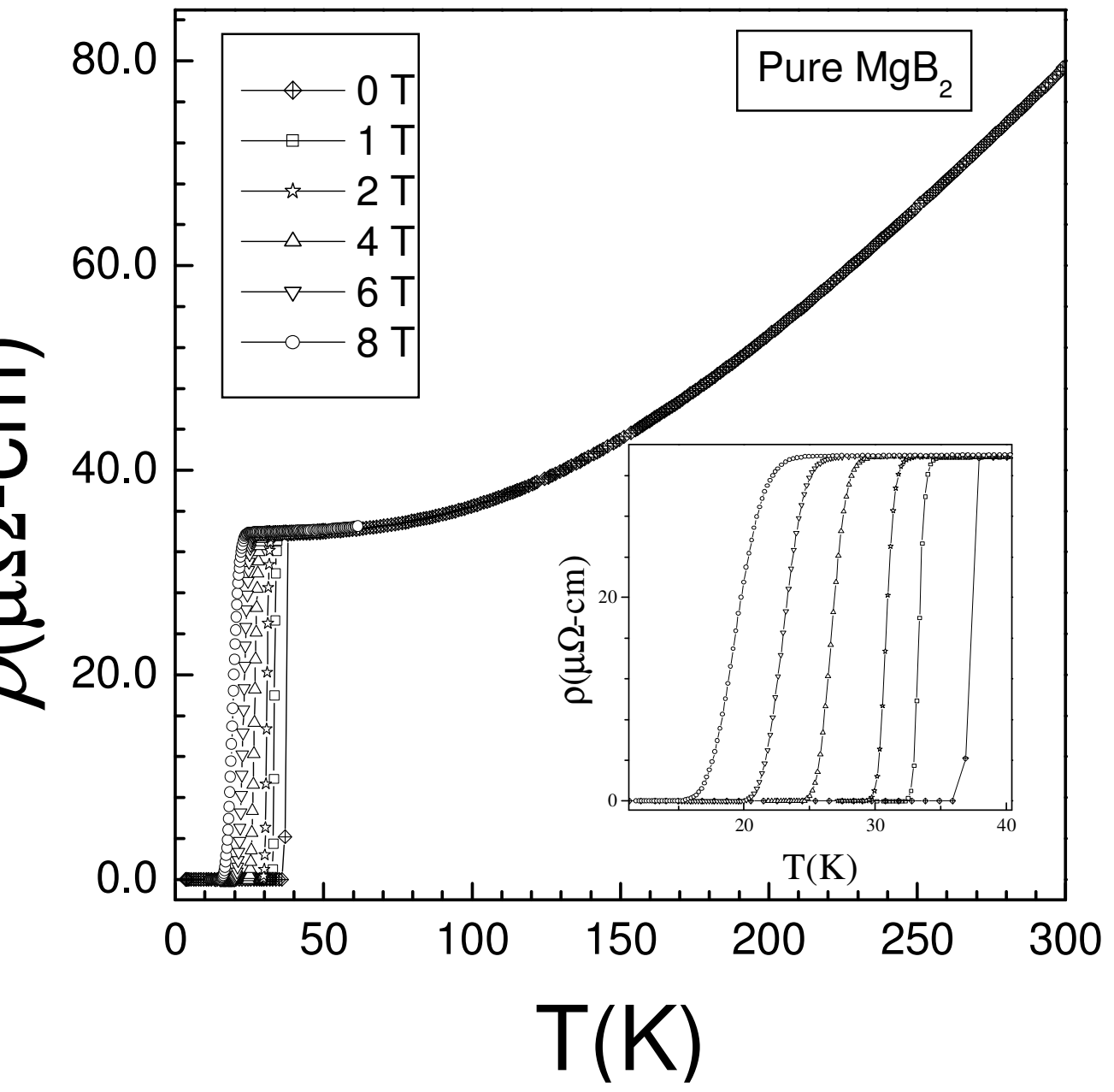

(a)

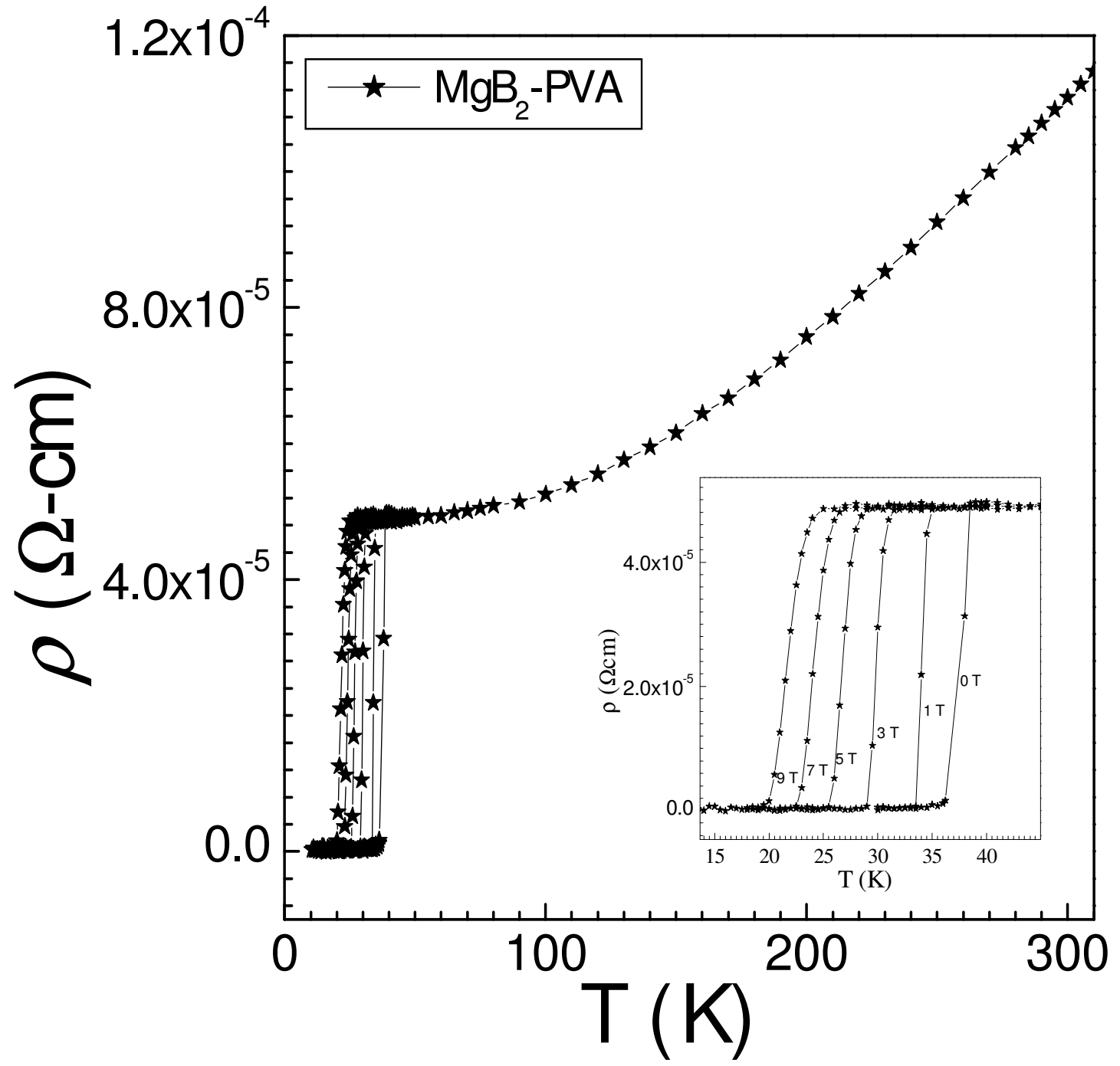

(b)

Fig. 3

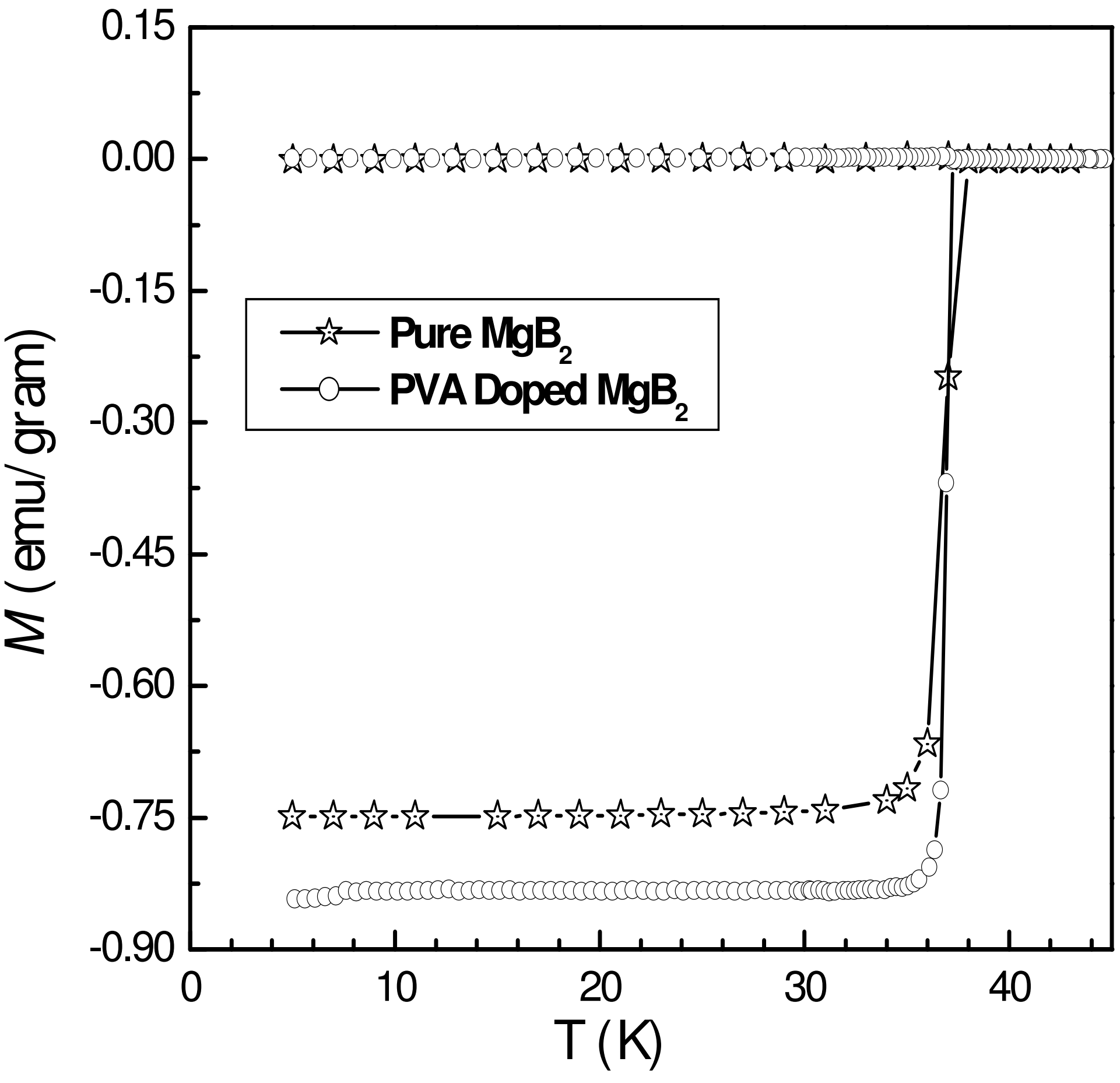

Fig. 4

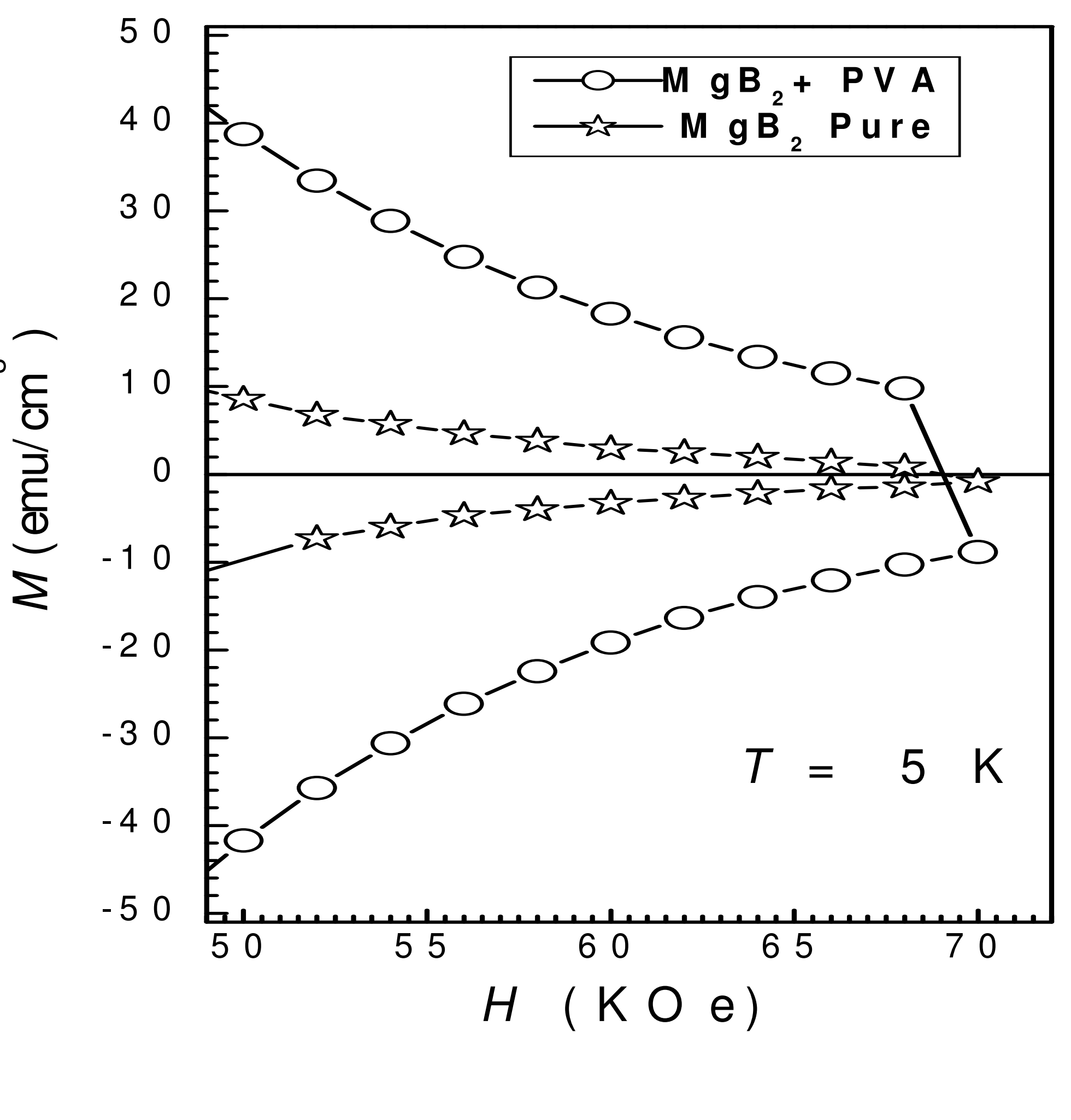

Fig. 5

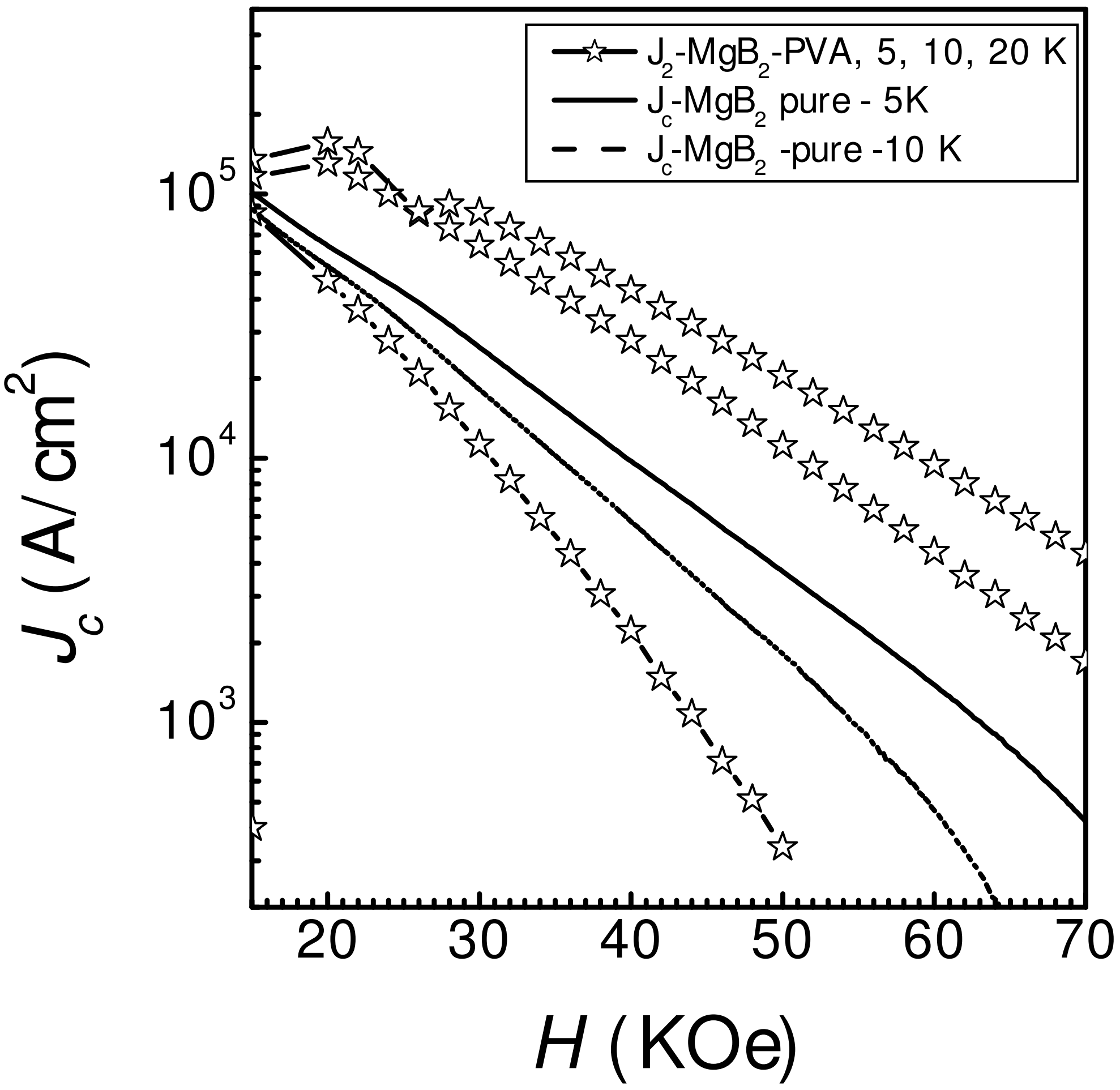

J2-MgB2-PVA, 5, 10, 20 K
Jc-MgB2 pure - 5K
Jc-MgB2 -pure -10 K
10^5
10^4
10^3
20
30
40
50
60
70
Jc (A/cm^2)
H (KOe)

Fig. 6

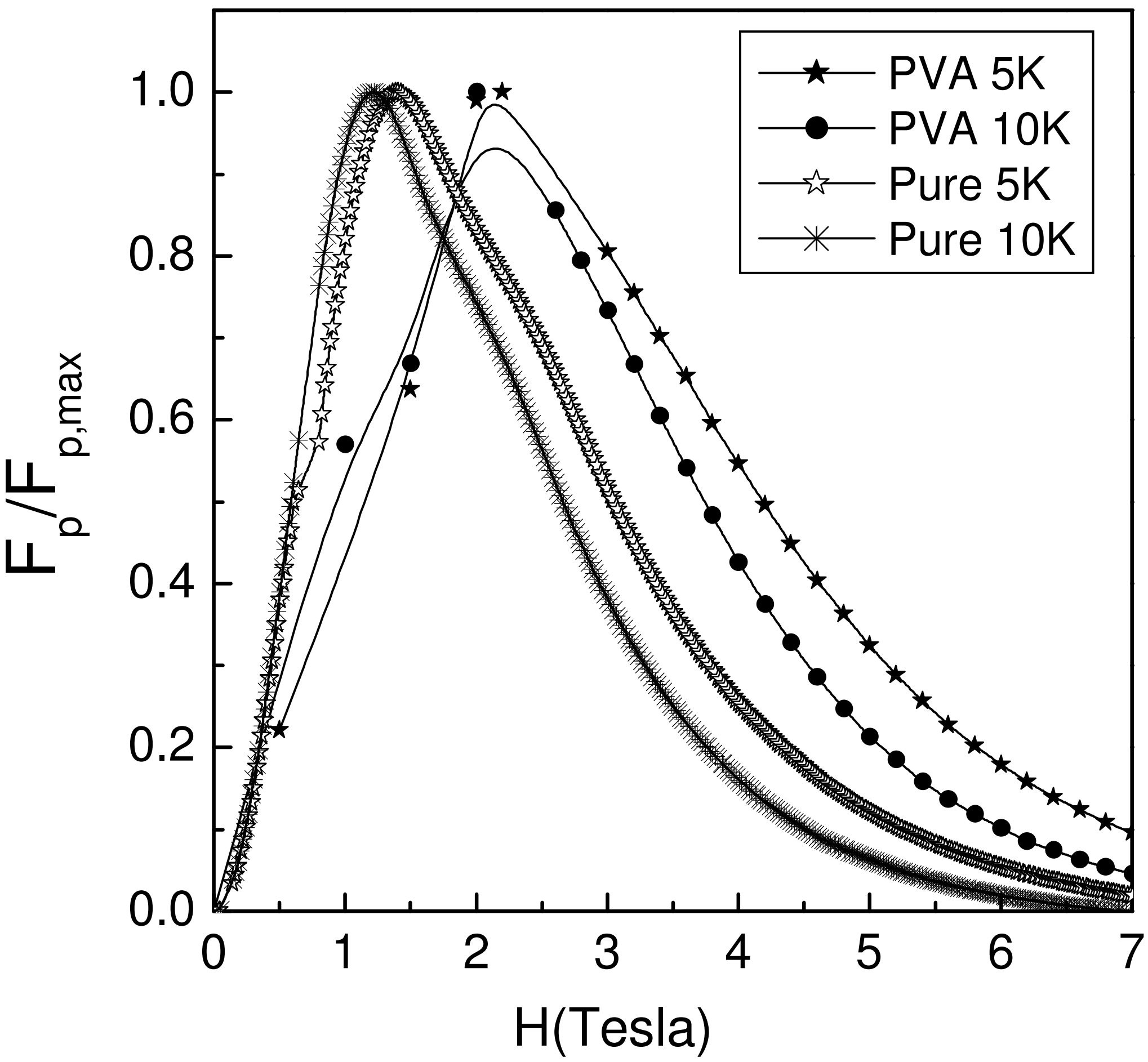

Fig. 7

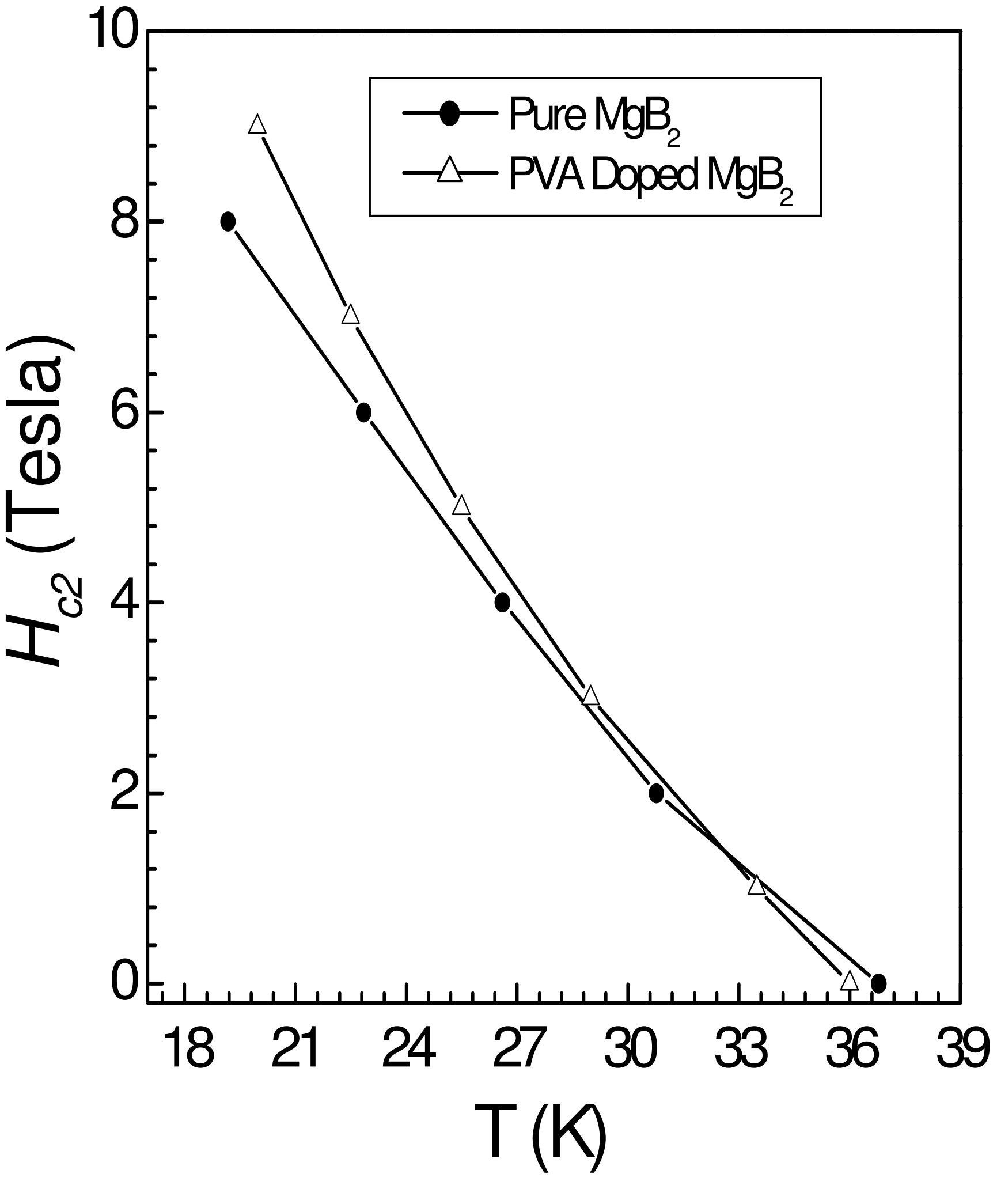